\documentclass{PoS}

\title{Multiwavelength studies of hard X-ray selected sources}

\ShortTitle{Multiwavelength studies of hard X-ray sources}

\author{\speaker{Pietro Parisi}\\
        INAF -- Istituto di Astrofisica Spaziale e Fisica Cosmica di Bologna, Via Gobetti 101, I-40129 Bologna, Italy.\\
        INAF -- Istituto di Astrofisica Spaziale e Fisica Cosmica di Roma, Via Fosso del Cavaliere  100, I-00133 Roma, Italy.\\
        E-mail: \email{parisi@iasf-roma.inaf.it}}
\author{on behalf of the IBIS survey team}

\abstract{Hard X-ray surveys like those provided by IBIS and BAT\
on board the INTEGRAL and Swift satellites list a significant\
number of sources which are unidentified and/or unclassified\
and which deserve multiwaveband observations to be properly\
characterized.\
In this work we have been able to follow up 148 such sources,\
finding 27 X-ray binaries and 121 Active Galactic Nuclei (AGN).\
From the AGN sample we extracted a set of 94 AGN, belonging\
to the INTEGRAL/IBIS and Swift/BAT surveys, for which we\
performed an X-ray study to determine absorption and 2-10 keV\
flux by means of XMM-Newton and Swift/XRT available observations.\
Using a new diagnostic diagram we identified a few peculiar\
sources which apparently do not fit within the AGN unified\
theory.\
Finally, we have compared the optical versus X-ray properties\
of these 94 AGN to study the optical reddening\
versus the X-ray absorption local to the AGN.}

\FullConference{The Extreme and Variable High Energy Sky - extremesky2011,\\
		September 19-23, 2011\\
		Chia Laguna (Cagliari), Italy}

\begin{document}

\section{Introduction}
A critically important region of the astrophysical spectrum is the hard X-ray band,
from 15 to 200 keV. In this
band, an unusually rich range of astrophysical processes occurs. 
This is the energy  domain where fundamental 
changes from primarily thermal to non-thermal sources/phenomena are 
expected, where the effects of absorption are drastically reduced and 
where most of the extreme astrophysical behaviour takes place.

Surprisingly, for such a pivotal region of the astronomical spectrum, the hard X-ray
Universe has been relatively
unexplored until recently. A number of surveys at energies higher than 10 keV are however now available to study the high energy sky. Indeed the surveys performed by IBIS on board
{\it INTEGRAL}, together with those made by BAT on board {\it Swift}, provide the best sample of objects selected in the hard X-ray band to date ([1], [2]).
They work in similar spectral bands, but concentrate on different parts of the sky: IBIS maps mainly the Galactic plane, while BAT focuses on observations at high Galactic latitudes.
All these pioneering studies indicated that the sky is rich of sources 
emitting above 10 keV, the majority being active galaxies of different types, X-ray binaries 
(with a black hole or a neutron star as accreting object) 
and pulsars. But there is more than producing a catalogue in the study of  the sky at energies
greater than 10 keV. 
Hard X-ray observations are crucial in studying absorbed sources and in exploring
spectral features.

In this work we have selected from the surveys of [2] for BAT and [1] and [4] for IBIS, those objects either without optical identification, or not well studied or without published optical spectra.
Following the method applied by [8], [9], [10], [11], [12] and [13] for the optical spectroscopic follow-up of unidentified {\it INTEGRAL} sources, 
we determine the nature of 148 selected objects, of which a large fraction ($\sim$82$\%$) turned out to be active galaxies. 
The remaining objects were found to be binaries of different types and will not be discussed
further here (for more details see Parisi 2011 Bologna University Ph.D. thesis).  

From the sample of AGN, we extracted a set of 94 objects for which we performed
an X-ray followup study by means of {\it XMM-Newton} and {\it Swift}/XRT observations. We
determined their X-ray spectral properties and in particular, for most objects in
the sample, we measured photon index, absorption column  density and X-ray flux.
Using a diagnostic diagram (see. [6]) we identified a few peculiar sources like unabsorbed
type 2 sources and absorbed type 1 objects: the first ones could be Compton thick AGN candidates (sources with column density greater than 10$^{24}$ cm$^{-2}$), but
also AGN without a Broad Line Region, while the second group may be composed of AGN viewed from a particular line of sight.
Both types of objects  are unexpected in the classical unified model of AGN.
Finally, in the last part we have compared the optical properties of
these 94 AGN with their X-ray characteristics and we discussed our overall results.

\section{Observations and data analysis}
For the optical follow up work, we used a set of 9 optical telescopes (2 national and 7 international) 
plus archival spectra from 6dF\footnote{{\tt http://www.aao.gov.au/local/www/6dF}} and SDSS\footnote{{\tt http://www.sdss.org}}.\\
The data reduction was performed with the standard procedure using IRAF as described in [8], [9], [10], [11], [12] and [13].

\begin{figure*}[th!]
\begin{center}
\includegraphics[width=4cm,angle=0]{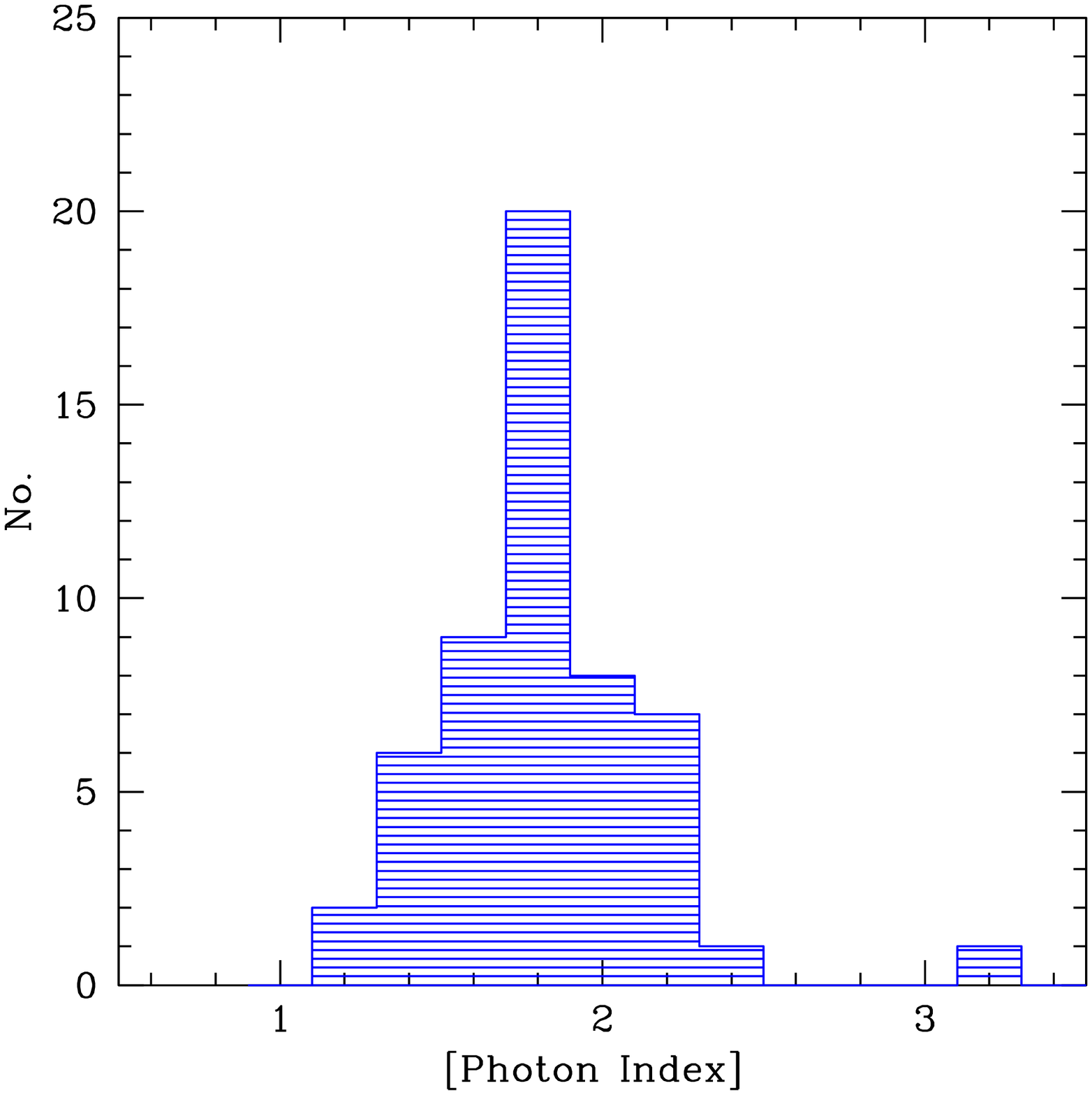}
\includegraphics[width=4cm,angle=0]{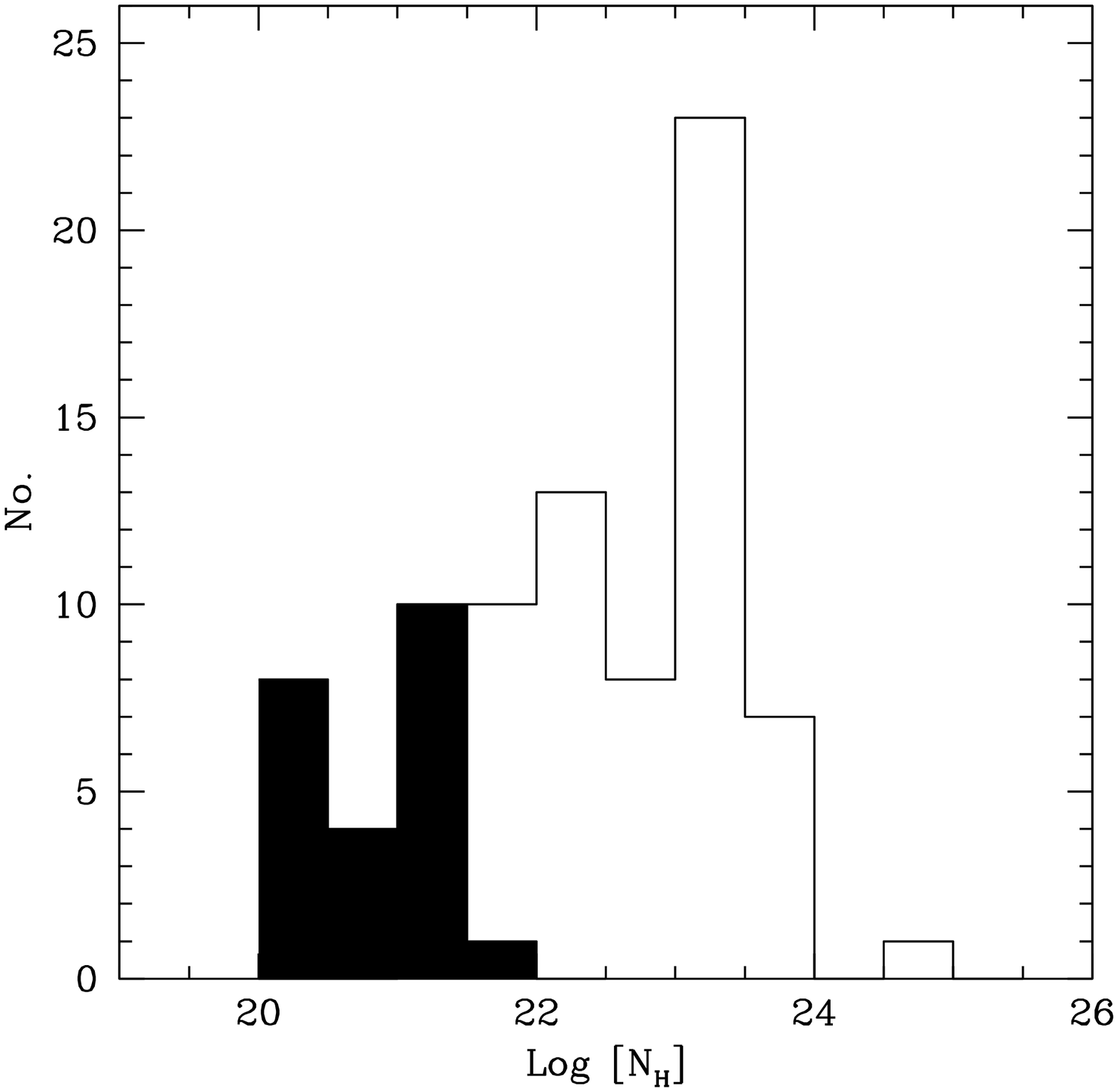}
\caption{On the left: distribution of the photon index of the AGN belonging to our sample and peaking at 1.83$\pm$0.34. On the right: 
column density logarithmic distribution for the entire sample of AGN, at an average value 22.6$\pm$0.8 (including upper limits). The black filled bins represent the sources for 
which only upper limits to N$_H$ are available.}\label{photon}
\end{center}
\end{figure*}

The X-ray data analysis was performed using Swift/XRT pointings for 85 out of 94 objects, while for the remaining ones archival XMM-\emph{Newton} 
observations have been considered. Due to the low statistics available, we chose for each source the best energy range for the spectral analysis and 
we employed as our baseline model,  a simple power law (often fixing the photon index to a canonical value of 1.8),
absorbed by  the Galactic ([3]) and, when required, the intrinsic neutral hydrogen column density. In figure \ref{photon} we report the 
photon index and the Log(N$_H$) distributions for the total sample composed of 94 AGN.
The average values for these distributions are respectively 1.83$\pm$0.34 and 22.6$\pm$0.8.

\section{X-ray versus optical classification}
The basic hypothesis of the unified theory of AGN is that the X-ray
absorption and optical obscuration are tightly related: absorbed AGN
should be classified as type 2 in optical (narrow permitted and forbidden lines are visible, so that in the optical spectra we only see the Narrow Line Region), 
while unabsorbed ones are expected to be of type 1 (broad permitted lines and narrow forbidden lines are visible, as we see both the Broad and the Narrow Line Regions). 

\begin{figure*}[th!]
\begin{center}
\includegraphics[width=7cm,angle=0]{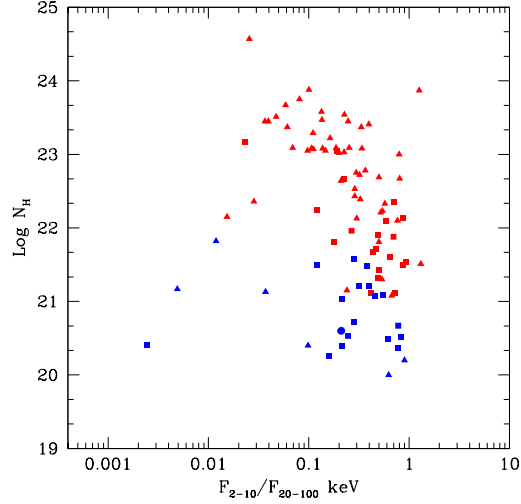}
\caption{Column density plotted against the F2-10 keV/F20-100 keV flux ratio of our sample of AGN. 
Squares are type 1 objects, triangles are type 2 AGN. The blue symbols represent those objects for which we only have galactic N$_H$, i.e.
where no intrinsic absorption has been measured.}\label{diagn}
\end{center}
\end{figure*}

This relationship is not always respected as it is evident in our sample (made of 38 type 1 and 56 type 2 objects\footnote{For type 1 objects we means Seyfert 1-1.9 according to the classification reported in [16]}) where we find absorption in 
a number of type 1 objects and no absorption in a number of type 2 sources. In particular, unabsorbed type 2 AGN, could be either unrecognized Compton thick, or heavily obscured objects, or alternatively "naked" type 2 sources, i.e. those lacking the broad line region.
To discriminate among these classes, we use the diagnostic diagram provided by [6]. This diagram uses the N$_H$ versus softness ratio
F$_{(2-10 keV)}$/F$_{(20-100 keV)}$ to look for Compton thick AGN candidates and its validity
has recently been confirmed by [15] and [7]: misclassified Compton thick objects populate the part
of the diagram with low absorption and low softness ratios. Figure
\ref{diagn} shows this diagnostic tool applied to our sample.
The 20-100 keV data have been obtained from the IBIS and BAT surveys ([1] and [2]).
In the case of BAT data we estimated 
the 20-100 keV flux from the published 14-195 keV value, assuming a power law of photon index $\Gamma$ = 2.02.
In the plot of figure \ref{diagn}, an indication
of decreasing softness ratio as the absorption increases is visible as
expected if the 2-10 keV flux is progressively depressed when the
absorption becomes stronger. Blue (red) symbols in the figure represent
upper limits (measured values) on the column density both for type 1 and type 2 objects, which are indicated with red squares and triangles respectively. As anticipated, most of our sources follow the expected
trend except for a few objects which have too low a softness ratio for
the observed column density, suggesting a Compton thick nature. 
We therefore suggest that 3 of our sources can be new Compton thick AGN candidates for which further analysis is necessary 
(i.e. those Seyfert 2 galaxies with F$_{(2-10 keV)}$/F$_{(20-100 keV)}$ below a value of 0.05), while 3 more can be possibly naked Seyfert 2 
(i.e. those Seyfert 2 galaxies with F$_{(2-10 keV)}$/F$_{(20-100 keV)}$ above a value of 0.1 and with only a galactic N$_H$);  lastly 7 are absorbed 
type 1 objects, where the intrinsic N$_H$ is greater than 10$^{22}$ cm$^{-2}$.\\
In any case further analysis is necessary because, a low softness ratio may also arise from source variability, since follow-up observations 
with XMM and Swift/XRT were not taken simultaneously with INTEGRAL.


\vspace{-0.5cm}

\section {Optical versus X-ray absorption}
\begin{figure*}[th!]
\begin{center}
\includegraphics[width=5cm,angle=0]{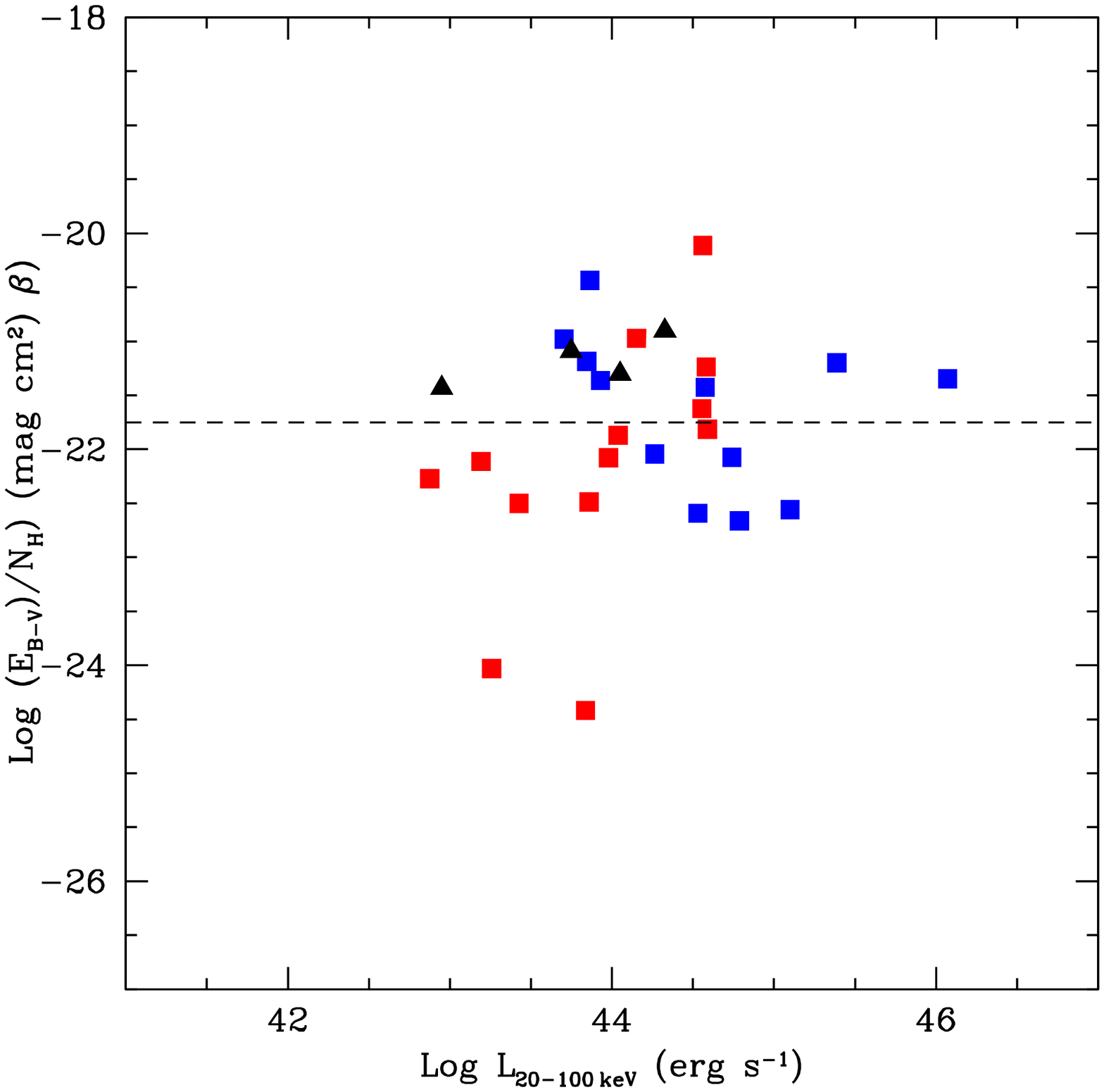}
\includegraphics[width=5cm,angle=0]{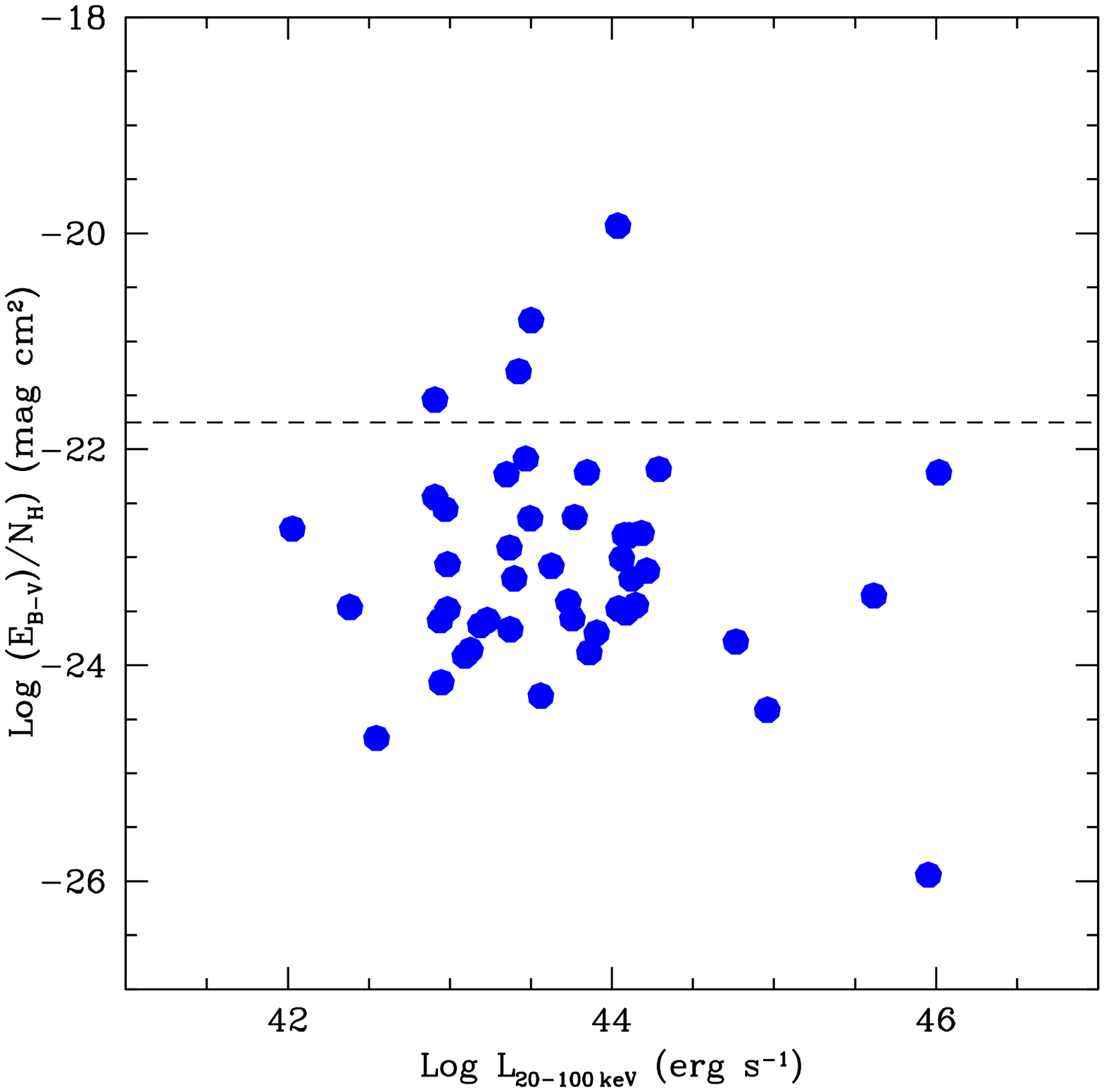}
\caption{On the left: E$_{B-V}$/N$_H$ ratio versus the 20-100 keV luminosity. The dashed line is the Galactic standard value. In the plot red squares show type 1.5 AGN, black triangles show type 1.8/1.9 AGN, while blue square type 1/1.2 AGN. On the right: E$_{B-V}$/N$_H$ ratio versus the 20-100 keV luminosity of type 2 AGN.}\label{Enh}
\end{center}
\end{figure*}

Various studies have shown that X-ray and optical nuclear absorptions do not always
match in AGNs. In particular, the measured optical extinction seems to be  
lower than that inferred from the absorption  measured in
X-rays, assuming a Galactic gas-to-dust ratio ([5], [16]); this evidence comes so far from  studies done on small samples of AGN.
This is an important effect since it could provide important observational 
consequences such as  a mismatch between optical and X-ray
classification. From our analysis, we found that the Balmer decrement and hence the reddening E(B-V) 
is lower in the broad line than in the narrow line sources; this is similar to what observed
in X-rays where type 1 objects tend to have lower absorbing column densities than type 2 AGN. 
Therefore we do not expect a big mismatch between the optical and the X-ray classification.
Clearly a one to one comparison is necessary and now possible with our sample of AGN. For this, we use the 
E(B-V) calculated from the Balmer decrement using the following formula derived from [14]: $$E(B-V)= a\; Log \left(\frac{H_{\alpha}/H_{\beta}}{(H_{\alpha}/H_{\beta})_0}\right),$$
where H$_{\alpha}/H_{\beta}$ is the observed Balmer decrement, $(H_{\alpha}/H_{\beta})_0$ is the intrinsic one (2.86), while
{\it a} is a constant (2.21), and the X-ray column density estimated from the X-ray spectra discussed 
in the previous chapter. Following [5], we choose to 
show in Figure \ref{Enh} the E$_{B-V}$/N$_H$ ratio as a function of the 20-100 keV luminosity; the dotted line reported in the figure represents the standard 
Galactic gas-to-dust ratio. It is worth noting that E(B-V) refers to extinction in the broad line region (BLR) in Seyfert 1s and to the narrow line region (NLR) in Seyfert 2;
it is therefore  not surprising that low E$_{B-V}$/N$_H$ values are found for Seyfert 2s, since narrow lines originate mostly
outside the region where the X-ray absorption takes place. Indeed we find that the  E$_{B-V}$/N$_H$ ratio is around the Galactic standard 
for the majority of type 1 sources (Fig. \ref{Enh}, right panel); the only two exceptions are  
IGR J12107+3822 and IGR J09253+6929, which have a very low E$_{B-V}$/N$_H$. 
Both have a conspicuous X-ray column density and are intermediate Seyferts. The first source 
has a good quality 2-10 keV spectrum and so we can assume that the value of the X-ray column density is well constrained; 
this could be an intermediate Seyfert with complex absorption
where we see the nucleus not directly, but at some inclination angle so that our line of sight intercepts thicker material from the torus.
The second source has a low quality X-ray spectrum with huge errors in the estimate of the N$_H$ and a fixed photon index; it is thus possible that, with better
statistics, its position in the graph shifts to a E$_{B-V}$/N$_H$ ratio closer to the dotted line. 
 
Concerning the more heavily absorbed X-ray type 2 sources (Fig. \ref{Enh}, left panel), we observe two typologies: 
those with an E$_{B-V}$/N$_H$ ratio not very different from the galactic value
and those substantially lower than the Galactic standard by a factor of 10-100. The first objects are somehow unexpected and could be AGN 
with extra dust absorption in the NLR. Further observations of these objects could be of some interest.

\section {Discussion and conclusion}
To conclude, in our sample, composed of 38 type 1 and 56 type 2 active galaxies, we find that
the majority of our sources follow the unified theory of AGN, with absorbed AGN classified as type 2 and unabsorbed AGN as type 1.
Some objects are however somewhat peculiar, as they do not follow this trend.
Using the diagnostic diagram of [6], we isolated these peculiar sources and discriminate among absorbed Seyfert 1, Compton thick sources or naked Seyfert 2,
finding 3 possible Compton thick sources, 3 naked type 2 and 7 absorbed type 1 AGN.
Concerning the hard X-ray and optical properties, if we compare our work with respect to the work of [5] and more recently that of [16], 
our results are somehow different. Most of our type 1 AGN show similar amount of reddening in optical and X-rays which indicates a similar region (torus/BLR)
for the dust and gas extinction. On the other hand not  all type 2 AGN, but only a fraction, show low  E$_{B-V}$/N$_H$ ratios as expected 
given that  the optical extinction originate outside the region where the X-ray absorption takes place.

\end{document}